\documentclass[iop]{emulateapj}
\usepackage{apjfonts}

\usepackage[tight]{subfigure}
\usepackage{amsmath}
\usepackage{graphicx}
\usepackage{wasysym}
\usepackage{color}
\usepackage{verbatim}

\newcommand{\msun}{\ensuremath{\mathrm{M}_\odot}}

\def\lesssim{\mathrel{\hbox{\rlap{\hbox{\lower4pt\hbox{$\sim$}}}\hbox{$<$}}}}

\def\gtrsim{\mathrel{\hbox{\rlap{\hbox{\lower4pt\hbox{$\sim$}}}\hbox{$>$}}}}

\shorttitle{Isotropic versus anisotropic CR driven winds}
\shortauthors{R. Pakmor et al.}

\begin{document}

\title{Galactic winds driven by isotropic and anisotropic cosmic ray diffusion in disk galaxies}

\author
{
 R.~Pakmor$^{1}$,
 C.~Pfrommer$^{1}$,
 C.~M.~Simpson$^{1}$
 and V.~Springel$^{1,2}$
}

\altaffiltext{1}
{Heidelberger Institut f\"{u}r Theoretische Studien, 
  Schloss-Wolfsbrunnenweg 35, 
  69118 Heidelberg, Germany
}

\altaffiltext{2}
{Zentrum f\"ur Astronomie der Universit\"at Heidelberg,
  Astronomisches Recheninstitut, M\"{o}nchhofstr. 12-14, 69120
  Heidelberg, Germany
}

\begin{abstract}
  The physics of cosmic rays (CR) is a promising candidate for
  explaining the driving of galactic winds and outflows.  Recent
  galaxy formation simulations have demonstrated the need for active CR
  transport either in the form of diffusion or streaming to
  successfully launch winds in galaxies. However, due to computational
  limitations, most previous simulations have modeled CR transport
  isotropically. Here, we discuss high resolution simulations of
  isolated disk galaxies in a $10^{11}\msun$ halo with the moving mesh
  code {\sc Arepo} that include injection of CRs from supernovae,
  advective transport, CR cooling, and CR transport
  through isotropic or anisotropic diffusion. We show that either mode
  of diffusion leads to the formation of strong bipolar
  outflows. However, they develop significantly later in the
  simulation with anisotropic diffusion compared to the simulation with
  isotropic diffusion.  Moreover, we find that isotropic diffusion
  allows most of the CRs to quickly diffuse out of the disk, while in the
  simulation with anisotropic diffusion, most CRs remain in the
  disk once the magnetic field becomes dominated by its azimuthal
  component, which occurs after $\sim 300\,{\rm Myrs}$. This has
  important consequences for the gas dynamics in the disk. In
  particular, we show that isotropic diffusion strongly suppresses the
  amplification of the magnetic field in the disk compared to
  anisotropic or no diffusion models. We therefore
  conclude that reliable simulations which include CR transport
  inevitably need to account for anisotropic diffusion.
\end{abstract}


\section{Introduction}
\label{sec:intro}

Understanding the formation and evolution of galaxies is one of the
most fascinating problems in modern cosmology. The observed galaxy
luminosity and H{\sc i}-mass functions have much shallower faint-end
slopes than predicted by $\Lambda$CDM models; this is locally known as
the `missing satellites problem' of the Milky Way, which should
contain many more dwarf-sized subhalos than observed \citep[see][ for
a review]{Kravtsov2010}. Recent cosmological simulations have
identified feedback in the form of galactic winds as being primarily
responsible for solving this problem as well as for obtaining
realistic rotation curves of disk galaxies and for enriching the
intergalactic medium with metals \citep{SchayeOWLS, Guedes2011,
  Puchwein2013, Marinacci2013, Hopkins2014FIRE, Vogelsberger2014,
  Schaye2015Eagle}.  However, galactic winds in these simulations are
often included as a phenomenological model. Their basic properties,
such as the wind velocity or mass loading, are tuned to match observed
global relations of galaxies. Ideally, we would like to directly
simulate the physics responsible for the wind driving, but the exact
nature of this physical origin is still strongly debated.

One popular idea is that momentum-driven winds could result from
radiation pressure acting on dust grains and atomic lines in dense
gas, imparting enough momentum to accelerate the gas, potentially
explaining strong outflows in starburst galaxies \citep{Murray2005,
  Thompson2005}.  However, direct radiation-hydrodynamics simulations
of the Rayleigh-Taylor instability \citep{Krumholz2012} and
large-scale galaxy models \citep{Rosdahl2015, Skinner2015} fail to see
strong, mass-loaded winds because the radiation is not sufficiently
trapped (as assumed in more simplified analytical calculations) but
instead generates a channel structure along which a substantial
fraction of the radiation is able to escape.

CRs in galaxies have also been proposed to drive outflows in a number
of theoretical works \citep{Ipavich1975, Breitschwerdt1991,
Zirakashvili1996, Ptuskin1997, Breitschwerdt2002, Socrates2008, Everett2008,
Everett2010, Samui2010, Dorfi2012} or by using three-dimensional
simulations of the ISM \citep{Hanasz2013, Girichidis2016}.  Polarized
radio observations of edge-on galaxies show poloidal field lines at
the disk-halo interface \citep[e.g.,][]{Tuellmann2000}. This argues
for a dynamical mechanism that is responsible for reorienting the
toroidal magnetic field in the disk and may be explained by a
CR-driven Parker instability \citep{Rodrigues2016}. Indeed, recent
hydrodynamical simulations of the formation and evolution of disk
galaxies have shown that CR pressure can drive strong bipolar outflows
in disk galaxies provided they are allowed to stream \citep{Uhlig2012,
  Ruszkowski2016} or diffuse \citep{Booth2013, Salem2014, Salem2014b}
relative to the rest frame of the gas. However, they also showed that
injecting CRs at supernova remnants and only accounting for their
advective transport is not sufficient for launching winds
\citep{CRPaper, Pfrommer2016}.

\begin{figure*}
  \centering
  \includegraphics[width=\linewidth]{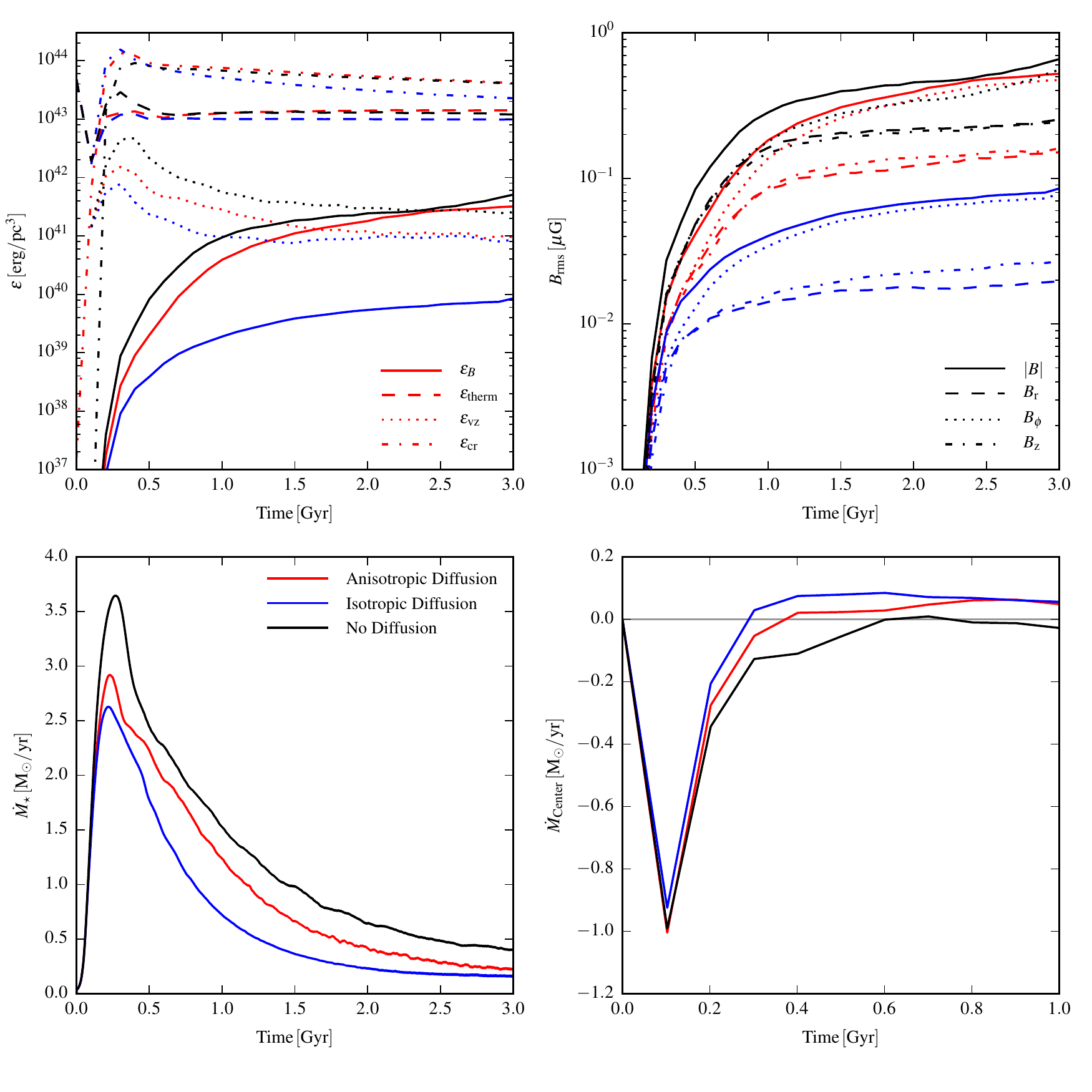}
  \caption{Evolution of different energy densities (top left),
      rms magnetic field strength (top right), star formation rate
      (bottom left), and the net vertical mass flux through a box of
      size 4~kpc centered on the galaxy (bottom right). Energies and
    magnetic field strength are measured in a cylindrical ring with
    inner radius $3$~kpc, outer radius $10$~kpc, and a height of
    $1$~kpc centered on the midplane. The inner radius is chosen to
    cut out the strong central bipolar outflow. The top left shows
    the evolution of the average magnetic energy density, thermal
    energy density, kinetic energy density in vertical velocities, and
    CR energy density, respectively. The top right panel shows
    $\mathrm{B_{rms}}$ for the total magnetic field strength and
    the radial, azimuthal, and vertical components in cylindrical
    coordinates, respectively. The bottom right panel is only shown until $t=1$~Gyr to
    emphasize the early evolution.}
  \label{fig:energies}
\end{figure*}

Most of these simulations treated CR transport in the isotropic
approximation, even though it is clear that CRs are dominantly
transported along magnetic field lines \citep{Desiati2014}.  Here, we
present high resolution simulations of the formation and evolution of
a disk galaxy in an isolated $10^{11}\msun$ dark matter halo with
isotropic as well as anisotropic diffusion and analyze the
differences. We describe our methods and setup in
\S\ref{sec:methods}. We then discuss differences in the properties of
the gas disk and the outflow in \S\ref{sec:disk}. We analyze the
magnetic field amplification in the disk for the different runs in
\S\ref{sec:bfield} and end with a brief summary in
\S\ref{sec:discussion}.

\newpage

\section{Methods and setup}
\label{sec:methods}

We simulate the formation and evolution of an isolated disk galaxy in
a $10^{11}\msun$ halo with the moving-mesh code \textsc{AREPO}
\citep{Arepo}. We use the new second order hydro scheme
\citep{Pakmor2016}. Cooling and star formation are modelled as
described in \citet{Springel2003}. Magnetic fields are modelled with
ideal MHD using cell-centered magnetic fields and the Powell scheme
\citep{Powell1999} for divergence control
\citep{Pakmor2011,Pakmor2013}.

CRs are modelled as a relativistic fluid with a constant
adiabatic index of $4/3$ in the two fluid approximation
\citep{Pfrommer2016} \footnote{Unlike \citet{Pfrommer2016} Sec.~3.3, here we use the collisional heating rate due to
Coulomb interactions only, where $\Gamma_\mathrm{th} = - \Lambda_{\mathrm{Coul}} = \tilde\lambda_\mathrm{th} n_\mathrm{e} \varepsilon_\mathrm{cr}$ and
$\tilde \lambda_\mathrm{th} =2.78 \times 10^{-16}~\mathrm{cm}^3~\mathrm{s}^{-1}$.}. To model the generation of CRs in
supernova remnants from core-collapse supernovae, we inject $10^{48}$
erg of CR energy per solar mass of formed stars into the local
environment of every newly created star particle. In contrast to
similar work \citep{Booth2013, Salem2014}, we describe the thermal and kinetic energy injection
with an effective equation of state \citep{Springel2003}. We model CR
cooling to thermal energy and radiation via Coulomb and hadronic
interactions assuming an equilibrium distribution
\citep{Pfrommer2016}. Moreover, in addition to advection of the cosmic
ray fluid with the gas we include active CR transport relative
to the gas rest frame in the form of isotropic or anisotropic
diffusion \citep{Pakmor2016Diffusion}. For isotropic diffusion, we
assume a constant isotropic diffusion coefficient of
$10^{28} \mathrm{cm^{2}\, s^{-1}}$.  For anisotropic diffusion, we
employ the same constant value parallel to the
magnetic field and no diffusion perpendicular to it.

Our setup is very similar to the isolated disk setup described in
\citet{Pakmor2013}. We model the dark matter halo as a static NFW
\citep{NFW1} density profile with a concentration parameter of
$7.2$. In this dark matter halo we put slowly rotating gas
in hydrostatic equilibrium with a baryonic mass fraction
of $0.17$ and add rotation with a spin parameter of
$\lambda=0.05$. The magnetic field is initialized as a uniform
homogeneous seed field along the $x$-axis with an initial field
strength of $10^{-10}\,\mathrm{G}$. There are no CRs in the
initial setup.

We start with about $10^7$ gas cells with a target mass of
$1.5 \times 10^3\, \msun$, the same as the typical mass of a star
particle. Using explicit refinement and de-refinement we enforce that
the mass of all cells is within a factor of two of the target
mass. Moreover, we require that adjacent cells differ by less than a
factor of $10$ in volume, otherwise the larger cell is also
refined. We employ the standard mesh regularization scheme in
\textsc{AREPO} \citep{Vogelsberger2012,Pakmor2016}. In the following,
we discuss $3$ simulations with CRs that differ by either
employing transport with isotropic diffusion, by using anisotropic
diffusion, or by not including transport at all. Otherwise the
simulations are identical.

\section{Disk properties}
\label{sec:disk}

\begin{figure*}
  \centering
  \includegraphics[width=\linewidth]{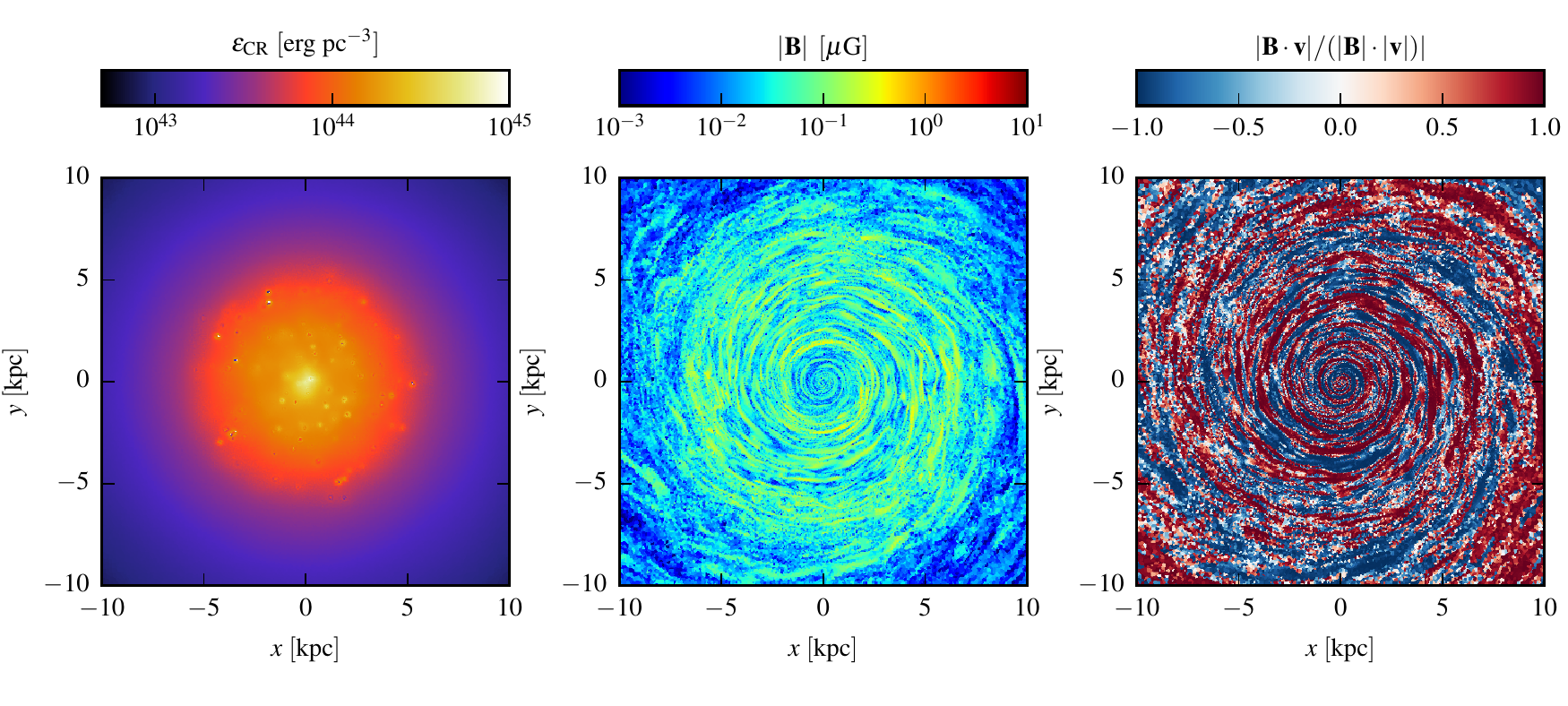}
  \includegraphics[width=\linewidth]{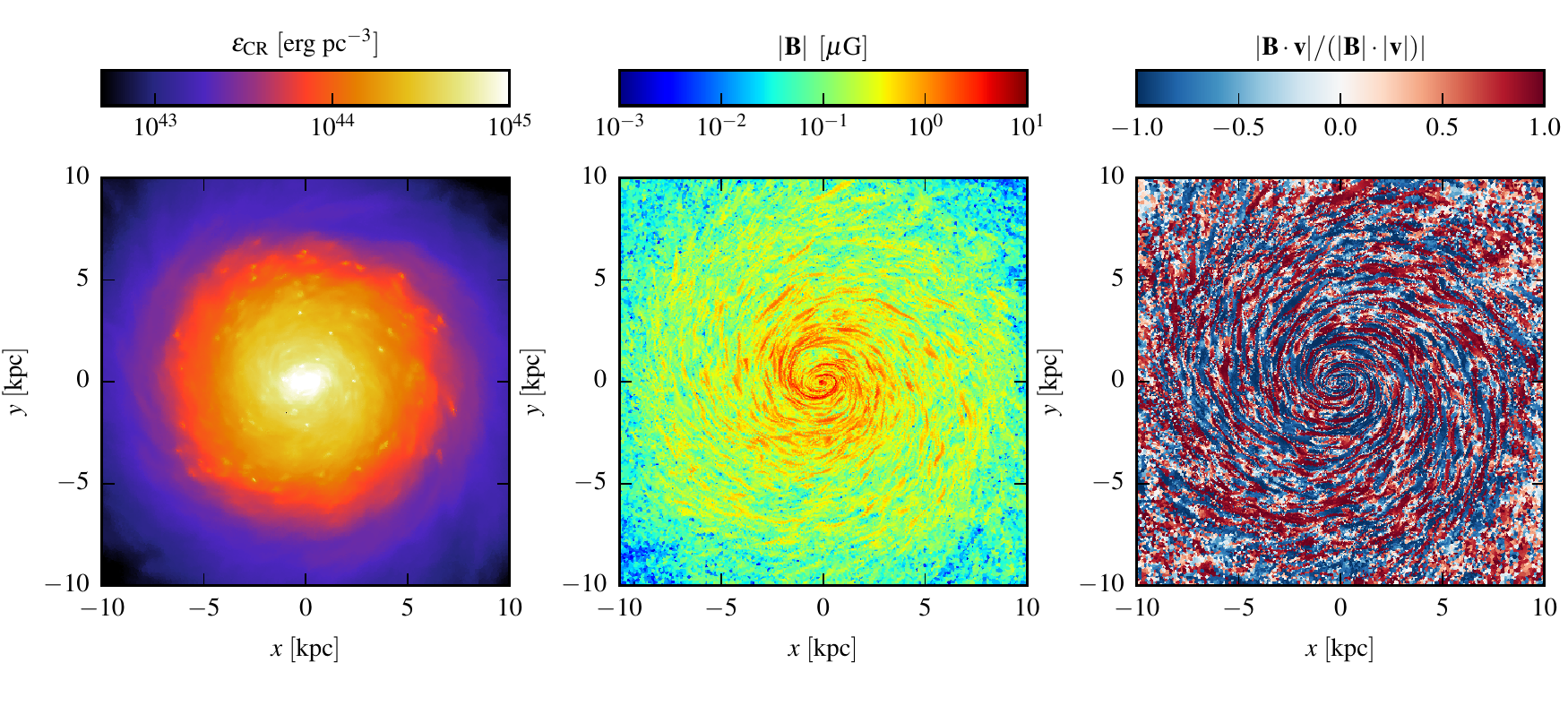}
  \caption{Properties of the gas disk after $1.5$~Gyrs for the runs
    with isotropic CR diffusion (top row) and anisotropic CR diffusion
    (bottom row). The columns show slices in the midplane of the
    disk. From left to right, the color-coded maps show the CR energy
    density, the magnetic field strength, and the cosine of the angle
    between the directions of the magnetic field and the velocity
    field.}
  \label{fig:diskslice}
\end{figure*}

\begin{figure*}
  \centering
  \includegraphics[width=\linewidth]{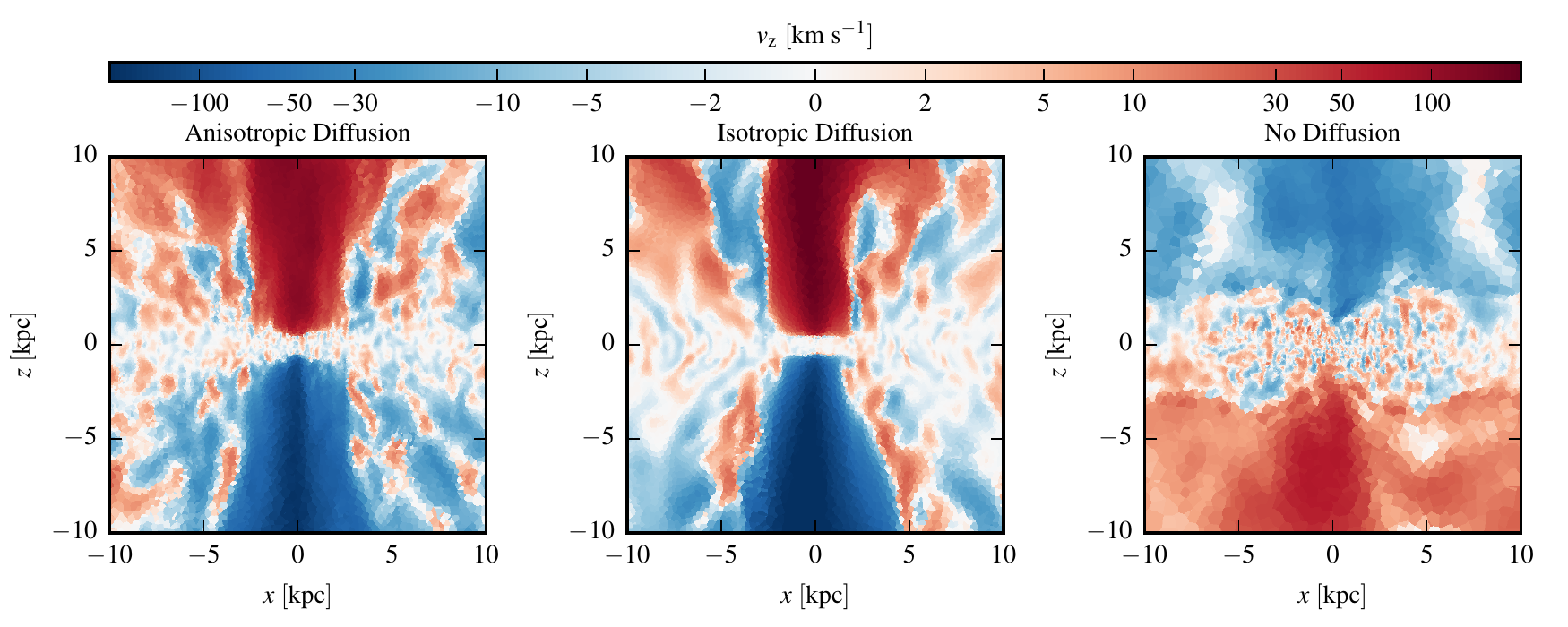}
  \caption{Slices in the $x-z$ plane through the center of the disk of
    the, showing the $z$-velocity component after $1.5$~Gyr. The
    columns from left to right correspond to the simulations with
    anisotropic CR diffusion, isotropic CR diffusion, and without CR
    diffusion, respectively.}
  \label{fig:wind}
\end{figure*}

Initially in all of our simulations, the gas sphere immediately
begins to cool, with the fastest cooling rates occurring at the dense
center. The gas loses pressure support and collapses into a disk as it
keeps its specific angular momentum. Once the gas density in the disk
becomes high enough, stars are formed and CRs are injected for
every star particle formed. Due to the resulting inhomogeneous cosmic
ray distribution, the transport of CRs becomes important. 

The star formation rate for all runs peaks
around $300$~Myr and then declines quickly, with most stars being
formed in the first Gyr (see Figure~\ref{fig:energies}). As the first
stars form, the CR energy density in the disk quickly
increases and is about $10$ times as large as the thermal energy
density when it peaks around $300$~Myr. Later it declines again,
mostly owing to cooling processes.

For our runs the star formation rate decreases with increasingly
efficient diffusion. This effect is caused by CRs diffusing away from the center 
of the disk where most stars are formed. They reduce the flow of gas to the 
center as shown in Fig.~\ref{fig:energies}. This is different to \citet{Salem2014}
who found that diffusion increases the star formation
rate. The discrepancy is likely a consequence of the different setup, as 
\citet{Salem2014} simulate the evolution of an existing disk which
leads to much more distributed star formation.

Figure~\ref{fig:diskslice} shows the disk after $1.5$~Gyr for the
runs with isotropic and anisotropic diffusion. At this time, the gas
densities in the disks are very similar in both runs. They strongly
peak in the center of the disks, where most of the stars are formed
and thus most of the CRs are injected. For the run with
isotropic diffusion, they are then transported away in all directions
and become quickly diluted into an almost spherical distribution. In
contrast, with anisotropic diffusion, CRs can only be
transported along magnetic field lines and therefore the diffusion process is
very sensitive to the structure of the magnetic field.

The magnetic field strength is initially (in the first $300$~Myr)
amplified exponentially on very small timescales to a field strength
of about $10^{-2}\,\mathrm{\mu G}$ (see top right panel in
Figure~\ref{fig:energies}), consistent with a turbulent small-scale
dynamo.  After this initial phase the disk has formed and the
differential rotation in the disk dominates the velocity field. The
magnetic field continues to grow exponentially, but on much longer
timescales, indicating that the dominant amplification mechanism has
changed. Around that time, the structure of the magnetic field also
changes from a chaotic small-scale field to one completely dominated
by its azimuthal component (see right panel in Figure
\ref{fig:diskslice}). Note that even though the magnetic field is
essentially completely aligned with the velocity field, it features a
large number of field reversals, a behavior that is clearly different
to our previous results for more massive halos \citep{Pakmor2013}. We
will explore this interesting dependence of the magnetic field
structure on halo mass in a separate publication.

For the simulation with anisotropic diffusion, this magnetic field structure means that within
the first $300$~Myrs, the CRs diffuse essentially
isotropically, but with a significantly lower effective diffusion
coefficient as they are not transported along straight lines on large
scales. Afterwards, however, they are mostly trapped at their radial
and vertical positions and are only carried along the angular
coordinate. Thus, a much larger fraction of the injected CRs
remains in the disk as compared to the run with isotropic
diffusion. Also, the run with anisotropic diffusion is more similar to
the simulation without any diffusion in terms of the overall
population of CRs in the disk.

Perhaps the strongest indication of a difference in the dynamic state
of the gas in the disk for the different types of diffusion can be
seen in the strength of the magnetic field. After $1.5$~Gyrs, the
typical magnetic field strength in the anisotropic diffusion run is
about $1\,\mathrm{\mu G}$.  In contrast, in the isotropic diffusion
run it only reaches a field strength of order $0.1\,\mathrm{\mu G}$ by
this time. Since the structure of the magnetic field is very similar
in both runs, it appears that either the same amplification mechanism
works with different efficiencies in the two runs, or that different
amplification mechanisms dominate in the two runs.

Even though the total energy in CRs in the disk is very
similar for the runs with anisotropic diffusion and without diffusion,
the effect of the CRs is rather different with respect to
outflows. As shown in Figure~\ref{fig:wind}, both simulations with
diffusion develop strong centralized bipolar outflows, comparable to
previous simulations of isolated disks with isotropic diffusion
\citep{Salem2014, Booth2013}. The wind is driven by the
vertical CR pressure gradient in the outer layers of the disk. At early
times $t<200$~Myr, its effect is very similar for both diffusion runs
(see Fig.~\ref{fig:energies}). Later, the magnetic field becomes dominated 
by its azimuthal component, and vertical CR diffusion is suppressed by 
a factor of $\approx 10$ for the anisotropic diffusion run. Its outflow fully 
develops after 800 Myr; by contrast, only 400 Myr are needed in the isotropic 
diffusion run.

After $1.5$~Gyr, the outflows have mass loading factors of $1.0$ 
(measured at a height of $5$~kpc above
and below the disk and within a radius of $10$~kpc) and $1.1$ in the
runs with anisotropic diffusion and isotropic diffusion, respectively,
broadly consistent with previous results for Milky Way like massive
halos \citep{Salem2014,Booth2013}.

\begin{figure*}
  \centering
  \includegraphics[width=\linewidth]{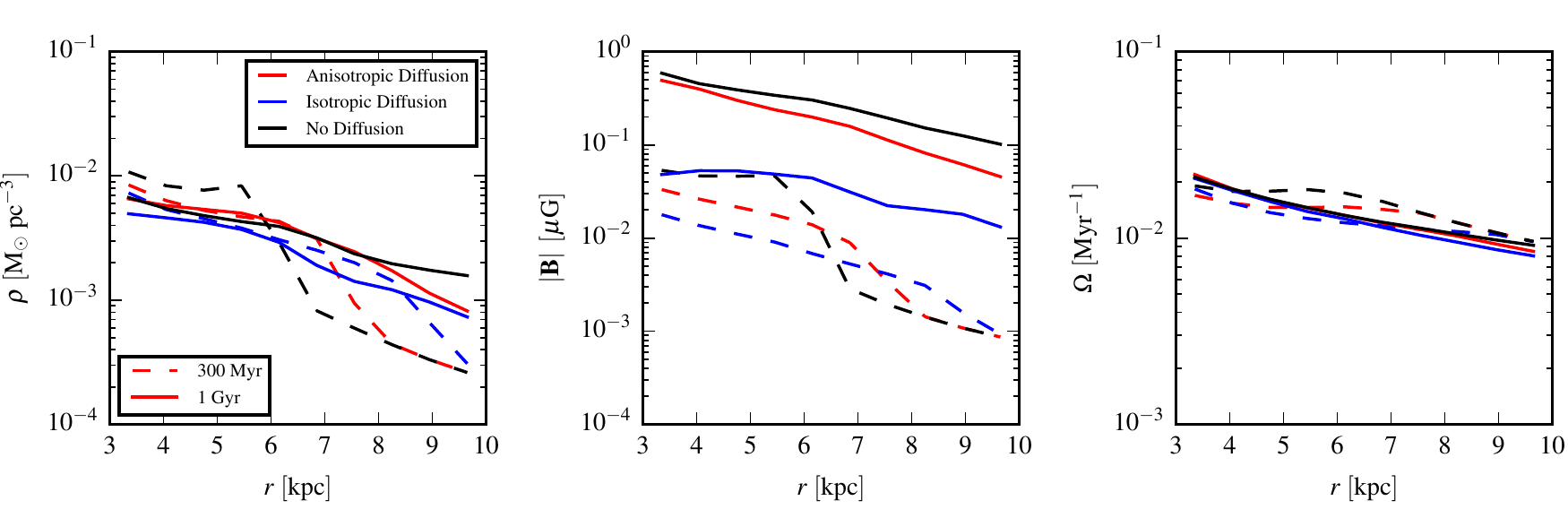}
  \caption{Radial profiles of volume weighted mean gas density, root mean square magnetic
    field strength, and angular velocity in a cylinder with a height
    of $8$~kpc centered on the midplane of the disk for the runs with
    anisotropic diffusion (black), isotropic diffusion (red), and
    without diffusion (blue), for two different times corresponding to
    $300$~Myrs (dashed lines) and $1.0$~Gyrs (solid lines),
    respectively.}
  \label{fig:rprof}
\end{figure*}

\begin{figure*}
  \centering
  \includegraphics[width=\linewidth]{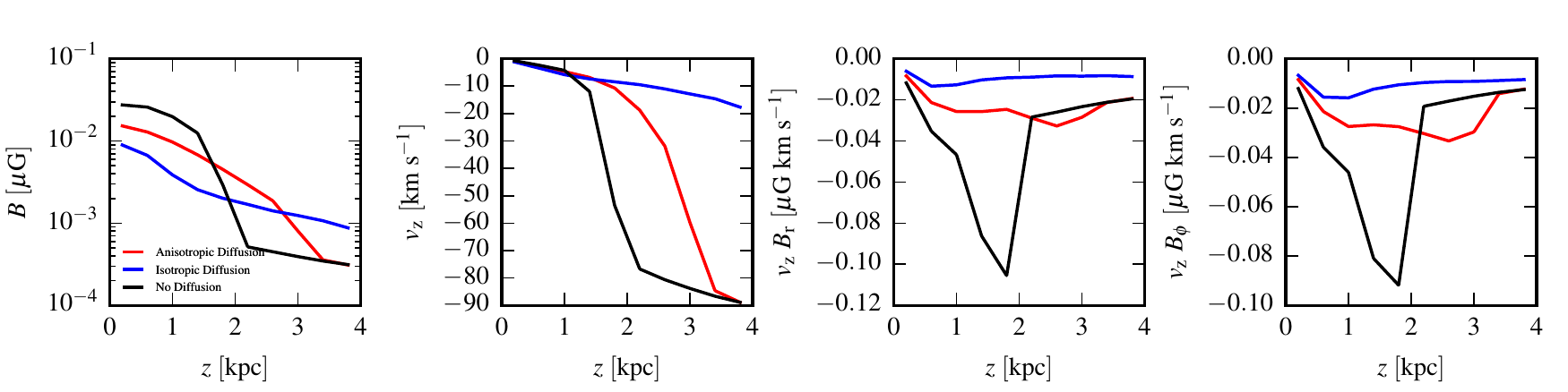}
  \includegraphics[width=\linewidth]{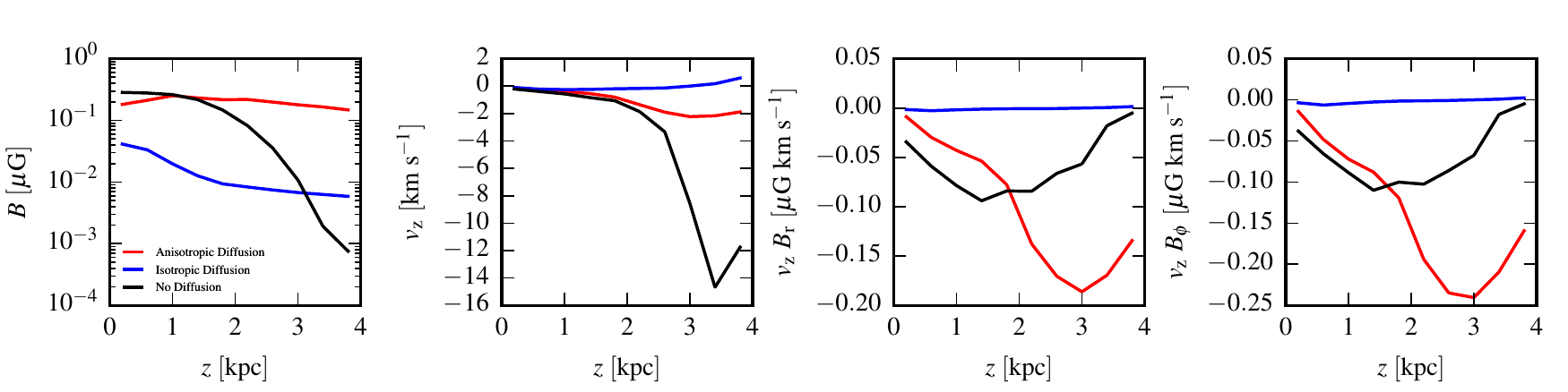}
  \caption{Vertical profiles of volume weighted mean gas properties in a cylindrical ring
    with an inner radius of $3$~kpc and an outer radius of $10$~kpc,
    after $300$~Myrs (top row) and $1.0$~Gyrs (bottom row), for the
    runs with anisotropic diffusion (black), isotropic diffusion
    (red), and without diffusion (blue). From left to right, the
    columns show total magnetic field strength, vertical velocity,
    vertical velocity times radial magnetic field strength, and
    vertical velocity times azimuthal magnetic field strength,
    respectively.}
  \label{fig:zprof}
\end{figure*}

The run without diffusion does not develop any large
scale outflows. Instead, it shows a strong fountain flow up to a
height of $2$-$3$~kpc. This fountain flow is also present in the
diffusion runs, but is suppressed in the simulation with isotropic
diffusion. In this calculation, the vertical velocity field in the
disk (excluding the bipolar outflow) seems to be generally lower and
lacks small-scale structure. Note that in both diffusion runs we
additionally see what looks like a large-scale fountain flow over the
disk, outside the central outflow, that reaches beyond $10$~kpc in
height.

\section{Magnetic field amplification}
\label{sec:bfield}

As discussed above, the amplification timescales of the magnetic field
strength deviate after an initial period of small-scale turbulent
amplification in the two runs with isotropic and anisotropic
diffusion. At the same time, the azimuthal component of the magnetic
field starts to dominate over the vertical and radial components in
all runs, even though initially all three components of the magnetic
field were of equal strength. This indicates a change in the dominant
amplification mechanism from small-scale turbulent amplification to a
large-scale dynamo.

An analytical model to describe magnetic field amplification in an
axisymmetric disk has been proposed by \citet{Shukurov2006}. In this
model, the mean radial field $\bar{B}_r$ and mean azimuthal field
$\bar{B}_\phi$ in cylindrical coordinates change as
\begin{eqnarray}
\frac{\partial \bar{B}_r}{\partial t}      = - \frac{\partial}{\partial z} \left( \bar{v}_z \bar{B}_r  + \mathcal{E}_\phi \right) ,\\
\frac{\partial \bar{B}_\phi}{\partial t} = - \frac{\partial}{\partial z} \left( \bar{v}_z \bar{B}_\phi  + \mathcal{E}_r \right) + q \Omega_0 \bar{B}_r,
\end{eqnarray}
where $\bar{v}_z$ is the mean vertical velocity in the disk,
$\mathcal{E}_r$ and $\mathcal{E}_\phi$ are the radial and azimuthal
components of the turbulent electric field, and $\Omega_0$ and $q$
parameterize the angular velocity as
$\Omega \left( r \right) = \Omega_0 \, r^q$. Here, we already
neglected additional terms containing ohmic diffusion, as assumed by
ideal MHD in our simulations.

The angular velocity shown in Figure~\ref{fig:rprof} can be well
described by a power law that it is very similar for all
runs. Therefore, it cannot explain the differences in the observed
amplification timescales of the magnetic field. Note that the
magnetic field strength already differs by a factor of two at
$300$~Myrs between the runs with isotropic and anisotropic diffusion.

To estimate the importance of the remaining terms in the galactic
dynamo equations above, we show the vertical profiles of the total
magnetic field strength, mean vertical velocity, and
$\bar{v}_z \bar{B}_r$ and $\bar{v}_z \bar{B}_\phi$ in
Figure~\ref{fig:zprof}. At $300$~Myrs, when the magnetic field
strength starts to diverge, the vertical velocities close to the disk
are still very similar in all three runs. However, the gradients in
the strength of the radial and vertical magnetic field are shallower
for the isotropic diffusion run as some highly magnetized gas has been
pushed vertically out of the disk. Therefore, the amplification terms
are smallest for the run with isotropic diffusion. The largest
amplification terms are present for the run without diffusion. It also
has the strongest gradients in the magnetic field strength as there
is no significantly magnetized material above a height of $2$~kpc. 

The situation is very similar after $1.5$~Gyrs. Now the magnetic field
strength has mostly saturated for the runs without diffusion and with
anisotropic diffusion, as the magnetic energy has become comparable to
the kinetic energy in the vertical component of the velocity
field. The still ongoing amplification is sufficient to compensate
losses by numerical diffusion of the magnetic field and by the
transport of magnetized gas out of the disk. In the run with isotropic
diffusion the magnetic field strength is one order of magnitude
smaller. It still increases slowly, but steadily, until its energy
approaches the vertical kinetic energy after about $10$~Gyrs.

Note that even though the galactic dynamo model seems to do a good job
in explaining the amplification of the magnetic field strength in our
simulations as well as the qualitative differences between the runs,
it is important to point out that its applicability is not obvious, as
the large-scale magnetic field has many local field reversals that
make the definition and interpretation of a mean magnetic field
difficult. We also note that we do not model ohmic diffusion,
therefore all magnetic diffusion is due to numerical dissipation
effects on the grid scale.

\section{Summary}
\label{sec:discussion}

In this work, we focussed on the impact of different CR
transport processes in simulations of isolated disk galaxies. For
clarity, we studied the same initial conditions with
isotropic diffusion and with anisotropic diffusion based on our newly
developed solver. In addition, we also applied to this set-up a model that does not explicitly 
account for diffusion processes.

Our simulation outcomes with isotropic diffusion are consistent with
previous results. However, we find that with anisotropic diffusion,
CRs are initially mostly confined to the disk due to the
predominantly azimuthal magnetic field, such that they end up being
distributed similarly to the run without any diffusion. Moreover, we
find that there are significant secondary differences in the disk gas
dynamics between the runs with isotropic diffusion and
anisotropic diffusion. In particular, the isotropic diffusion run
strongly suppresses the magnetic field amplification and needs almost
a Hubble time until the magnetic field strength saturates. In
comparison, in the simulations with anisotropic diffusion and without
diffusion, the field already saturates after around $1$~Gyr. 

Given these differences, we conclude that it is important to model the
full anisotropic nature of CR transport, because the isotropic
approximation can produce qualitatively and quantitatively incorrect
results. It also seems likely that a similar conclusion holds for the
more general case of CR streaming.  This process has been
included in simulations of galaxy formation for the first time by
\citet{Uhlig2012}, but in an isotropic fashion. Very recently,
\citet{Ruszkowski2016} presented the first attempt of an anisotropic
treatment. As our results confirm, this is clearly a very interesting
research direction for future work on galaxy formation and evolution.

\begin{acknowledgements}

This work has been supported by the European Research Council under
ERC-StG grant EXAGAL-308037, ERC-CoG grant CRAGSMAN-646955, and by the
Klaus Tschira Foundation. VS acknowledges support through subproject
EXAMAG of the Priority Programme 1648 ``Software for Exascale
Computing'' of the German Science Foundation.

\end{acknowledgements}

\bibliographystyle{apj}

\begin{thebibliography}{45}
\expandafter\ifx\csname natexlab\endcsname\relax\def\natexlab#1{#1}\fi

\bibitem[{{Booth} {et~al.}(2013){Booth}, {Agertz}, {Kravtsov}, \&
  {Gnedin}}]{Booth2013}
{Booth}, C.~M., {Agertz}, O., {Kravtsov}, A.~V., \& {Gnedin}, N.~Y. 2013,
  \apjl, 777, L16

\bibitem[{{Breitschwerdt} {et~al.}(2002){Breitschwerdt}, {Dogiel}, \&
  {V{\"o}lk}}]{Breitschwerdt2002}
{Breitschwerdt}, D., {Dogiel}, V.~A., \& {V{\"o}lk}, H.~J. 2002, \aap, 385, 216

\bibitem[{{Breitschwerdt} {et~al.}(1991){Breitschwerdt}, {McKenzie}, \&
  {Voelk}}]{Breitschwerdt1991}
{Breitschwerdt}, D., {McKenzie}, J.~F., \& {Voelk}, H.~J. 1991, \aap, 245, 79

\bibitem[{{Desiati} \& {Zweibel}(2014)}]{Desiati2014}
{Desiati}, P., \& {Zweibel}, E.~G. 2014, \apj, 791, 51

\bibitem[{{Dorfi} \& {Breitschwerdt}(2012)}]{Dorfi2012}
{Dorfi}, E.~A., \& {Breitschwerdt}, D. 2012, \aap, 540, A77

\bibitem[{{Everett} {et~al.}(2010){Everett}, {Schiller}, \&
  {Zweibel}}]{Everett2010}
{Everett}, J.~E., {Schiller}, Q.~G., \& {Zweibel}, E.~G. 2010, \apj, 711, 13

\bibitem[{{Everett} {et~al.}(2008){Everett}, {Zweibel}, {Benjamin}, {McCammon},
  {Rocks}, \& {Gallagher}}]{Everett2008}
{Everett}, J.~E., {Zweibel}, E.~G., {Benjamin}, R.~A., {et~al.} 2008, \apj,
  674, 258

\bibitem[{{Girichidis} {et~al.}(2016){Girichidis}, {Naab}, {Walch}, {Hanasz},
  {Mac Low}, {Ostriker}, {Gatto}, {Peters}, {W{\"u}nsch}, {Glover}, {Klessen},
  {Clark}, \& {Baczynski}}]{Girichidis2016}
{Girichidis}, P., {Naab}, T., {Walch}, S., {et~al.} 2016, \apjl, 816, L19

\bibitem[{{Guedes} {et~al.}(2011){Guedes}, {Callegari}, {Madau}, \&
  {Mayer}}]{Guedes2011}
{Guedes}, J., {Callegari}, S., {Madau}, P., \& {Mayer}, L. 2011, \apj, 742, 76

\bibitem[{{Hanasz} {et~al.}(2013){Hanasz}, {Lesch}, {Naab}, {Gawryszczak},
  {Kowalik}, \& {W{\'o}lta{\'n}ski}}]{Hanasz2013}
{Hanasz}, M., {Lesch}, H., {Naab}, T., {et~al.} 2013, \apjl, 777, L38

\bibitem[{{Hopkins} {et~al.}(2014){Hopkins}, {Kere{\v s}}, {O{\~n}orbe},
  {Faucher-Gigu{\`e}re}, {Quataert}, {Murray}, \& {Bullock}}]{Hopkins2014FIRE}
{Hopkins}, P.~F., {Kere{\v s}}, D., {O{\~n}orbe}, J., {et~al.} 2014, \mnras,
  445, 581

\bibitem[{{Ipavich}(1975)}]{Ipavich1975}
{Ipavich}, F.~M. 1975, \apj, 196, 107

\bibitem[{{Jubelgas} {et~al.}(2008){Jubelgas}, {Springel}, {Ensslin}, \&
  {Pfrommer}}]{CRPaper}
{Jubelgas}, M., {Springel}, V., {Ensslin}, T., \& {Pfrommer}, C. 2008, A\&A,
  481, 33

\bibitem[{{Kravtsov}(2010)}]{Kravtsov2010}
{Kravtsov}, A. 2010, Advances in Astronomy, 2010, 281913

\bibitem[{{Krumholz} \& {Thompson}(2012)}]{Krumholz2012}
{Krumholz}, M.~R., \& {Thompson}, T.~A. 2012, \apj, 760, 155

\bibitem[{{Marinacci} {et~al.}(2013){Marinacci}, {Pakmor}, \&
  {Springel}}]{Marinacci2013}
{Marinacci}, F., {Pakmor}, R., \& {Springel}, V. 2013, \mnras

\bibitem[{{Murray} {et~al.}(2005){Murray}, {Quataert}, \&
  {Thompson}}]{Murray2005}
{Murray}, N., {Quataert}, E., \& {Thompson}, T.~A. 2005, \apj, 618, 569

\bibitem[{{Navarro} {et~al.}(1996){Navarro}, {Frenk}, \& {White}}]{NFW1}
{Navarro}, J.~F., {Frenk}, C.~S., \& {White}, S.~D.~M. 1996, ApJ, 462, 563

\bibitem[{{Pakmor} {et~al.}(2011){Pakmor}, {Bauer}, \& {Springel}}]{Pakmor2011}
{Pakmor}, R., {Bauer}, A., \& {Springel}, V. 2011, \mnras, 418, 1392

\bibitem[{{Pakmor} {et~al.}(2016{\natexlab{a}}){Pakmor}, {Pfrommer}, {Simpson},
  {Kannan}, \& {Springel}}]{Pakmor2016Diffusion}
{Pakmor}, R., {Pfrommer}, C., {Simpson}, C.~M., {Kannan}, R., \& {Springel}, V.
  2016{\natexlab{a}}, ArXiv e-prints

\bibitem[{{Pakmor} \& {Springel}(2013)}]{Pakmor2013}
{Pakmor}, R., \& {Springel}, V. 2013, \mnras, 432, 176

\bibitem[{{Pakmor} {et~al.}(2016{\natexlab{b}}){Pakmor}, {Springel}, {Bauer},
  {Mocz}, {Munoz}, {Ohlmann}, {Schaal}, \& {Zhu}}]{Pakmor2016}
{Pakmor}, R., {Springel}, V., {Bauer}, A., {et~al.} 2016{\natexlab{b}}, \mnras,
  455, 1134

\bibitem[{{Pfrommer} {et~al.}(2016){Pfrommer}, {Pakmor}, {Schaal}, {Simpson},
  \& {Springel}}]{Pfrommer2016}
{Pfrommer}, C., {Pakmor}, R., {Schaal}, K., {Simpson}, C.~M., \& {Springel}, V.
  2016, ArXiv e-prints

\bibitem[{{Powell} {et~al.}(1999){Powell}, {Roe}, {Linde}, {Gombosi}, \& {De
  Zeeuw}}]{Powell1999}
{Powell}, K.~G., {Roe}, P.~L., {Linde}, T.~J., {Gombosi}, T.~I., \& {De Zeeuw},
  D.~L. 1999, Journal of Computational Physics, 154, 284

\bibitem[{{Ptuskin} {et~al.}(1997){Ptuskin}, {Voelk}, {Zirakashvili}, \&
  {Breitschwerdt}}]{Ptuskin1997}
{Ptuskin}, V.~S., {Voelk}, H.~J., {Zirakashvili}, V.~N., \& {Breitschwerdt}, D.
  1997, \aap, 321, 434

\bibitem[{{Puchwein} \& {Springel}(2013)}]{Puchwein2013}
{Puchwein}, E., \& {Springel}, V. 2013, \mnras, 428, 2966

\bibitem[{{Rodrigues} {et~al.}(2016){Rodrigues}, {Sarson}, {Shukurov},
  {Bushby}, \& {Fletcher}}]{Rodrigues2016}
{Rodrigues}, L.~F.~S., {Sarson}, G.~R., {Shukurov}, A., {Bushby}, P.~J., \&
  {Fletcher}, A. 2016, \apj, 816, 2

\bibitem[{{Rosdahl} {et~al.}(2015){Rosdahl}, {Schaye}, {Teyssier}, \&
  {Agertz}}]{Rosdahl2015}
{Rosdahl}, J., {Schaye}, J., {Teyssier}, R., \& {Agertz}, O. 2015, \mnras, 451,
  34

\bibitem[{{Ruszkowski} {et~al.}(2016){Ruszkowski}, {Yang}, \&
  {Zweibel}}]{Ruszkowski2016}
{Ruszkowski}, M., {Yang}, H.-Y.~K., \& {Zweibel}, E. 2016, ArXiv e-prints,
  1602.04856

\bibitem[{{Salem} \& {Bryan}(2014)}]{Salem2014}
{Salem}, M., \& {Bryan}, G.~L. 2014, \mnras, 437, 3312

\bibitem[{{Salem} {et~al.}(2014){Salem}, {Bryan}, \& {Hummels}}]{Salem2014b}
{Salem}, M., {Bryan}, G.~L., \& {Hummels}, C. 2014, \apjl, 797, L18

\bibitem[{{Samui} {et~al.}(2010){Samui}, {Subramanian}, \&
  {Srianand}}]{Samui2010}
{Samui}, S., {Subramanian}, K., \& {Srianand}, R. 2010, \mnras, 402, 2778

\bibitem[{{Schaye} {et~al.}(2010){Schaye}, {Dalla Vecchia}, {Booth}, {Wiersma},
  {Theuns}, {Haas}, {Bertone}, {Duffy}, {McCarthy}, \& {van de
  Voort}}]{SchayeOWLS}
{Schaye}, J., {Dalla Vecchia}, C., {Booth}, C.~M., {et~al.} 2010, \mnras, 402,
  1536

\bibitem[{{Schaye} {et~al.}(2015){Schaye}, {Crain}, {Bower}, {Furlong},
  {Schaller}, {Theuns}, {Dalla Vecchia}, {Frenk}, {McCarthy}, {Helly},
  {Jenkins}, {Rosas-Guevara}, {White}, {Baes}, {Booth}, {Camps}, {Navarro},
  {Qu}, {Rahmati}, {Sawala}, {Thomas}, \& {Trayford}}]{Schaye2015Eagle}
{Schaye}, J., {Crain}, R.~A., {Bower}, R.~G., {et~al.} 2015, \mnras, 446, 521

\bibitem[{{Shukurov} {et~al.}(2006){Shukurov}, {Sokoloff}, {Subramanian}, \&
  {Brandenburg}}]{Shukurov2006}
{Shukurov}, A., {Sokoloff}, D., {Subramanian}, K., \& {Brandenburg}, A. 2006,
  \aap, 448, L33

\bibitem[{{Skinner} \& {Ostriker}(2015)}]{Skinner2015}
{Skinner}, M.~A., \& {Ostriker}, E.~C. 2015, \apj, 809, 187

\bibitem[{{Socrates} {et~al.}(2008){Socrates}, {Davis}, \&
  {Ramirez-Ruiz}}]{Socrates2008}
{Socrates}, A., {Davis}, S.~W., \& {Ramirez-Ruiz}, E. 2008, \apj, 687, 202

\bibitem[{{Springel}(2010)}]{Arepo}
{Springel}, V. 2010, \mnras, 401, 791

\bibitem[{{Springel} \& {Hernquist}(2003)}]{Springel2003}
{Springel}, V., \& {Hernquist}, L. 2003, \mnras, 339, 289

\bibitem[{{Thompson} {et~al.}(2005){Thompson}, {Quataert}, \&
  {Murray}}]{Thompson2005}
{Thompson}, T.~A., {Quataert}, E., \& {Murray}, N. 2005, \apj, 630, 167

\bibitem[{{T{\"u}llmann} {et~al.}(2000){T{\"u}llmann}, {Dettmar}, {Soida},
  {Urbanik}, \& {Rossa}}]{Tuellmann2000}
{T{\"u}llmann}, R., {Dettmar}, R.-J., {Soida}, M., {Urbanik}, M., \& {Rossa},
  J. 2000, \aap, 364, L36

\bibitem[{{Uhlig} {et~al.}(2012){Uhlig}, {Pfrommer}, {Sharma}, {Nath},
  {En{\ss}lin}, \& {Springel}}]{Uhlig2012}
{Uhlig}, M., {Pfrommer}, C., {Sharma}, M., {et~al.} 2012, \mnras, 423, 2374

\bibitem[{{Vogelsberger} {et~al.}(2012){Vogelsberger}, {Sijacki}, {Kere{\v s}},
  {Springel}, \& {Hernquist}}]{Vogelsberger2012}
{Vogelsberger}, M., {Sijacki}, D., {Kere{\v s}}, D., {Springel}, V., \&
  {Hernquist}, L. 2012, \mnras, 425, 3024

\bibitem[{{Vogelsberger} {et~al.}(2014){Vogelsberger}, {Genel}, {Springel},
  {Torrey}, {Sijacki}, {Xu}, {Snyder}, {Nelson}, \&
  {Hernquist}}]{Vogelsberger2014}
{Vogelsberger}, M., {Genel}, S., {Springel}, V., {et~al.} 2014, \mnras, 444,
  1518

\bibitem[{{Zirakashvili} {et~al.}(1996){Zirakashvili}, {Breitschwerdt},
  {Ptuskin}, \& {Voelk}}]{Zirakashvili1996}
{Zirakashvili}, V.~N., {Breitschwerdt}, D., {Ptuskin}, V.~S., \& {Voelk}, H.~J.
  1996, \aap, 311, 113

\end{thebibliography}

\end{document}